\documentclass[11pt,afpaper]{article}
\usepackage{jheppub}

\pdfoutput=1

\newcommand{\nn}{\nonumber}
\def\eqa{\begin{eqnarray}}
\def\eqae{\end{eqnarray}}
\def\eq{\begin{equation}}
\def\eqe{\end{equation}}
\def\be{\begin{equation}}
\def\ee{\end{equation}}
\def\bea{\begin{eqnarray}}
\def\eea{\end{eqnarray}}
\def\ba{\begin{array}}
\def\ea{\end{array}}
\def\bd{\begin{displaymath}}
\def\ed{\end{displaymath}}

\def\>{\rangle}
\def\<{\langle}

\newcommand{\fft}[2]{\frac{#1}{#2}}
\newcommand{\ft}[2]{{\textstyle\frac{#1}{#2}}}

\preprint{MCTP-17-08}

\title{Toward Microstate Counting Beyond Large $N$ in Localization and the Dual One-loop Quantum Supergravity}

\author[a]{James T. Liu,}
\author[a,b]{Leopoldo A. Pando Zayas,}
\author[a]{Vimal Rathee,}
\author[a]{and Wenli Zhao}

\affiliation[a]{Michigan Center for Theoretical Physics,  Randall Laboratory of Physics,\\
The University of Michigan,  Ann Arbor, MI 48109-1040, USA}
\affiliation[b]{The Abdus Salam International Centre for Theoretical Physics,\\
Strada Costiera 11, 34014 Trieste, Italy}

\abstract{The topologically twisted index for ABJM theory with gauge group $U(N)_k \times U(N)_{-k}$ has recently been shown, in the large-$N$ limit, to reproduce the Bekenstein-Hawking entropy of certain magnetically charged asymptotically AdS$_4$ black holes.  We numerically study the index beyond the large-$N$ limit and provide evidence that it contains a subleading logarithmic term of the form $-1/2\log N$.  On the holographic side, this term naturally arises from a one-loop computation.  However, we find that the contribution coming from the near horizon states does not reproduce the field theory answer.  We give some possible reasons for this apparent discrepancy.}

\begin{document}
\maketitle

\section{Introduction}

It is often said that black holes are the hydrogen atom of quantum gravity since they are systems where aspects of quantum gravity are indispensable for their understanding. If this analogy is to be pursued then studying corrections to the Bekenstein-Hawking entropy should be equivalent to going beyond the Bohr energy levels. History teaches us that the leading term can be obtained  without a full understanding of the degrees of freedom but that the corrections require a detailed control of various aspects of the system. In this work, we take some steps into the subleading corrections to the microstate counting of the entropy of a class of magnetically charged asymptotically AdS$_4$ black holes. 

Indeed, Strominger and Vafa have demonstrated that string theory provides a framework for the microstate counting of the Bekenstein-Hawking entropy of a class of asymptotically flat black holes \cite{Strominger:1996sh}. Moreover, Sen and collaborators have carried a successful program of understanding logarithmic corrections to various black holes   \cite{Banerjee:2010qc,Banerjee:2011jp,Sen:2011ba,Sen:2012cj}. 

In the context of the AdS/CFT correspondence \cite{Maldacena:1997re}, a microscopic counting of the Bekenstein-Hawking entropy of a class of black holes has recently been presented by Benini, Hristov and Zaffaroni \cite{Benini:2015eyy,Benini:2016rke}.  Understanding black hole entropy in this context is particularly powerful because it does provide a practical path to a fully non-perturbative definition of quantum gravity in asymptotically AdS spacetimes.  The basic premise of \cite{Benini:2015eyy} is that the topologically twisted index of ABJM, namely the supersymmetric partition function on $S^1\times S^2$ with background magnetic flux on $S^2$ \cite{Benini:2015noa}, counts the ground state degeneracy of a superconformal quantum mechanics on $S^1$, and that this counting enumerates the microstates of the dual magnetically charged BPS black hole in AdS.

It was in fact demonstrated in \cite{Benini:2015eyy} that the topologically twisted index reproduces the AdS black hole entropy at leading order in the large-$N$ expansion. Here, we wish to extend this correspondence to subleading order by examining the logarithmic corrections on both the field theory and gravity sides of the duality. 

In section~\ref{Sec:FieldTheory} we start by reviewing the field theory computation of the topologically twisted index and present numerical evidence pointing to a universal $-1/2 \log N$ correction.  We then turn to the gravity calculation in section~\ref{Sec:Gravity}, which first reviews the prescription and special status of logarithmic corrections at the one-loop level. We then discuss the dual calculation in the context of 11-dimensional supergravity, focusing on the contribution coming from the near horizon limit of the magnetically charged BPS black hole solutions. We also discuss the absence of potential contributions coming from the asymptotically AdS$_4$ region.  In contrast with the index result, we find $-2\log N$ from the quantum gravity computation, and suggest possible reasons for this discrepancy in section~\ref{Sec:Conclusions}.

\section{The topologically twisted index beyond the large-$N$ limit}\label{Sec:FieldTheory}

The topologically twisted index for three dimensional ${\cal N}=2$ field theories was defined in \cite{Benini:2015noa} (see other related work \cite{Honda:2015yha,Closset:2015rna,Hosseini:2016tor,Hosseini:2016ume,Closset:2016arn}) by evaluating the supersymmetric partition function on $S^1\times S^2$ with a topological twist on $S^2$.  When applied to the microstate counting of magnetic AdS$_4$ black holes, the index is computed for ABJM theory, and the topological twist arises from the magnetic fluxes on $S^2$ \cite{Benini:2015eyy,Benini:2016rke}.  Since these black holes are constructed in the STU model truncation of four-dimensional SO(8) gauged supergravity, there are a total of four U(1) gauge fields, with corresponding charges $n_a$ satisfying the supersymmetry constraint $\sum n_a=2$.

The topologically twisted index for ABMJ theory was worked out in \cite{Benini:2015eyy}, and reduces to the evaluation of
the partition function
\be
Z(y_a,n_a)=\prod_{a=1}^4 y_a^{-\frac{1}{2}N^2 n_a}\sum_{I\in BAE}\frac{1}{\det\mathbb{B}}
\frac{\prod_{i=1}^N x_i^N \tilde{x}_i^N\prod_{i\neq j}\left(1-\frac{x_i}{x_j}\right)\left(1-\frac{\tilde{x}_i}{\tilde{x}_j}\right)}{\prod_{i,j=1}^N\prod_{a=1,2}(\tilde{x}_j-y_ax_i)^{1-n_a}\prod_{a=3,4}(x_i-y_a\tilde{x}_j)^{1-n_a}},
\label{eq:logZ}
\ee
where $y_a$ are the corresponding fugacities.  The summation is over all solutions $I$ of the ``Bethe Ansatz Equations" (BAE) $e^{iB_i}=e^{i\tilde{B}_i}=1$ modulo permutations, where
\begin{align}
e^{iB_i}&=x_i^k\prod_{j=1}^N\frac{(1-y_3 \frac{\tilde{x}_j}{x_i})(1-y_4 \frac{\tilde{x}_j}{x_i})}{(1-y_1^{-1} \frac{\tilde{x}_j}{x_i})(1-y_2^{-1} \frac{\tilde{x}_j}{x_i})},\nn\\
e^{i\tilde{B}_j}&=\tilde{x}_j^k\prod_{i=1}^N\frac{(1-y_3 \frac{\tilde{x}_j}{x_i})(1-y_4 \frac{\tilde{x}_j}{x_i})}{(1-y_1^{-1} \frac{\tilde{x}_j}{x_i})(1-y_2^{-1} \frac{\tilde{x}_j}{x_i})}.
\end{align}
Here $k$ is the Chern-Simons level, and the two sets of variables $\{x_i\}$ and $\{\tilde x_j\}$ arise from the $U(N)_k\times U(N)_{-k}$ structure of ABJM theory.  Finally, the $2N\times 2N$ matrix $\mathbb{B}$ is the Jacobian relating the $\{x_i,\tilde x_j\}$ variables to the $\{e^{iB_i},e^{i\tilde B_j}\}$ variables
\begin{equation}
\mathbb B=\begin{pmatrix}x_l\fft{\partial e^{iB_j}}{\partial x_l}&\tilde x_l\fft{\partial e^{iB_j}}{\partial\tilde x_l}\\[4pt]
x_l\fft{\partial e^{i\tilde B_j}}{\partial x_l}&\tilde x_l\fft{\partial e^{i\tilde B_j}}{\partial\tilde x_l}\end{pmatrix}.
\end{equation}
See \cite{Benini:2015eyy} for additional details.

It is convenient to introduce the chemical potentials $\Delta_a$ according to $y_a=e^{i\Delta_a}$ and furthermore perform a change of variables $x_i=e^{iu_i}$, $\tilde{x}_j=e^{i\tilde{u}_j}$.  In this case, the BAE become
\begin{align}
0&=ku_i -i\sum_{j=1}^N\left[\sum_{a=3,4}\log\left(1-e^{i(\tilde{u}_j-u_i+\Delta_a)}\right)-
\sum_{a=1,2}\log\left(1-e^{i(\tilde{u}_j-u_i-\Delta_a)}\right)\right]-2\pi n_i, \nonumber \\
0&=k\tilde{u}_j -i\sum_{i=1}^N\left[\sum\limits_{a=3,4}\log\left(1-e^{i(\tilde{u}_j-u_i+\Delta_a)}\right)-
\sum_{a=1,2}\log\left(1-e^{i(\tilde{u}_j-u_i-\Delta_a)}\right)\right]-2\pi \tilde{n}_j.
\label{eq:BAE}
\end{align}
The topologically twisted index is evaluated by first solving these equations for $\{u_i,\tilde u_j\}$, and then inserting the resulting solution into the partition function (\ref{eq:logZ}).  This procedure was carried out in \cite{Benini:2015eyy} in the large-$N$ limit with $k=1$ by introducing the parametrization
\begin{equation}
u_i=iN^{1/2}\, t_i +\pi-\ft12\delta v(t_i), \qquad \tilde{u}_i=iN^{1/2}\,t_i+\pi+\ft12\delta v(t_i),
\end{equation}
where we have further made use of reflection symmetry about $\pi$ along the real axis.  In the large-$N$ limit, the eigenvalue distribution becomes continuous, and the set $\{t_i\}$ may be described by an eigenvalue density $\rho(t)$.

\subsection{Evaluation of the index beyond the leading order in $N$}

\begin{figure}[t]
\centering
\includegraphics[width=0.7\linewidth]{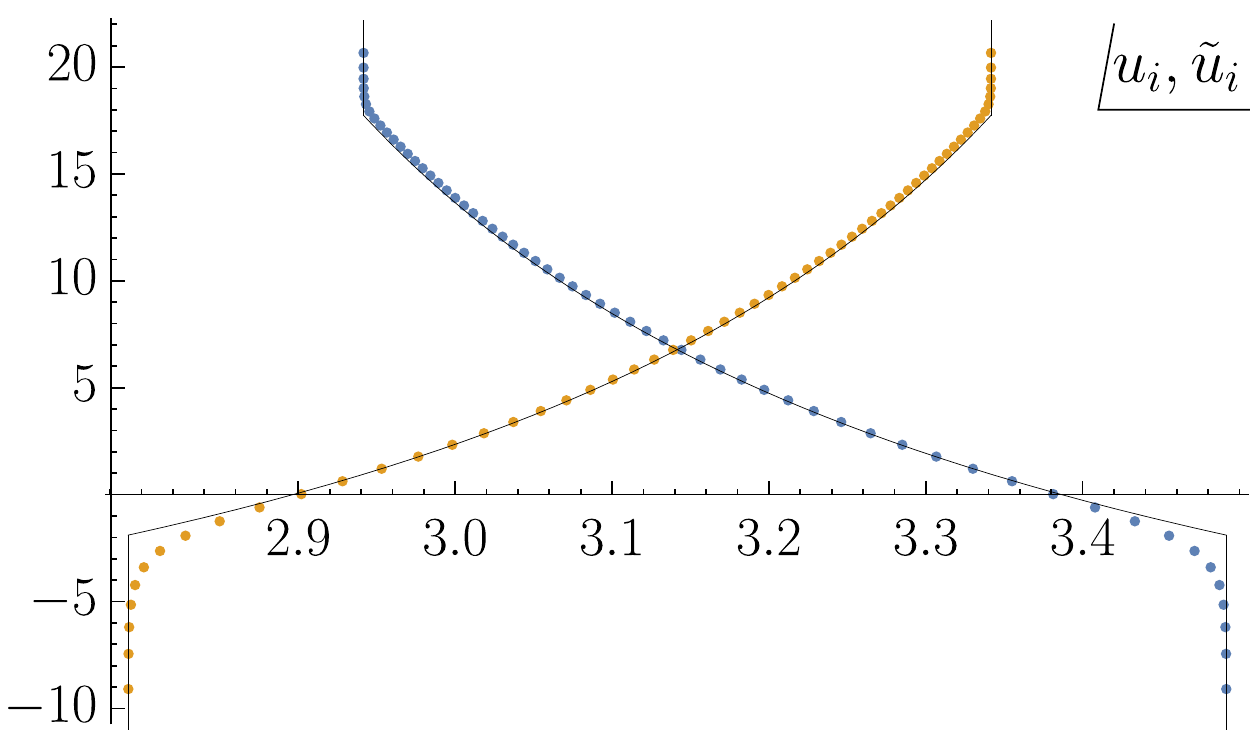}
\caption{The solution to the BAE for $\Delta_a=\{0.4,0.5,0.7,2\pi-1.6\}$ and $N=60$.  The solid lines correspond to the leading order expression obtained in \cite{Benini:2015eyy}.}
\label{fig:evals}
\end{figure}

The leading order solution for $\rho(t)$ and $\delta v(t)$ was worked out in \cite{Benini:2015eyy}, and the resulting partition function exhibits the expected $N^{3/2}$ scaling of ABJM theory
\be
{\rm Re}\log Z_0=-\frac{N^{3/2}}{3}\sqrt{2\Delta_1\Delta_2\Delta_3\Delta_4}\sum\limits_a\frac{n_a}{\Delta_a}.
\label{eq:logZ0}
\ee
A similar result was extended to the context of asymptotically AdS$_4$ black holes with hyperbolic horizon in \cite{Cabo-Bizet:2017jsl}.  We are, of course, interested in taking this solution beyond the leading order.  In the ABJM context, we expect the subleading behavior of the index to have the form
\begin{equation}
{\rm Re}\log Z={\rm Re}\log Z_0+N^{1/2}f_1(\Delta_a,n_a)+\log Nf_2(\Delta_a,n_a)+f_3(\Delta_a,n_a)+\mathcal O(N^{-1/2}),
\label{eq:ABJMrlz}
\end{equation}
where the functions $f_1$, $f_2$ and $f_3$ are linear in the magnetic fluxes $n_a$.  In principle, we would like to systematically extend the analysis beyond the leading order in order to obtain the analytic form of these functions.  However, this appears to be a challenge, mainly due to the presence of the (left and right) tails of the eigenvalue distribution. (These tails correspond to the nearly vertical segments in figure~\ref{fig:evals}.)  We thus proceed with a numerical investigation.

\begin{figure}[t]
\centering
\hbox to \linewidth{\includegraphics[width=0.48\linewidth]{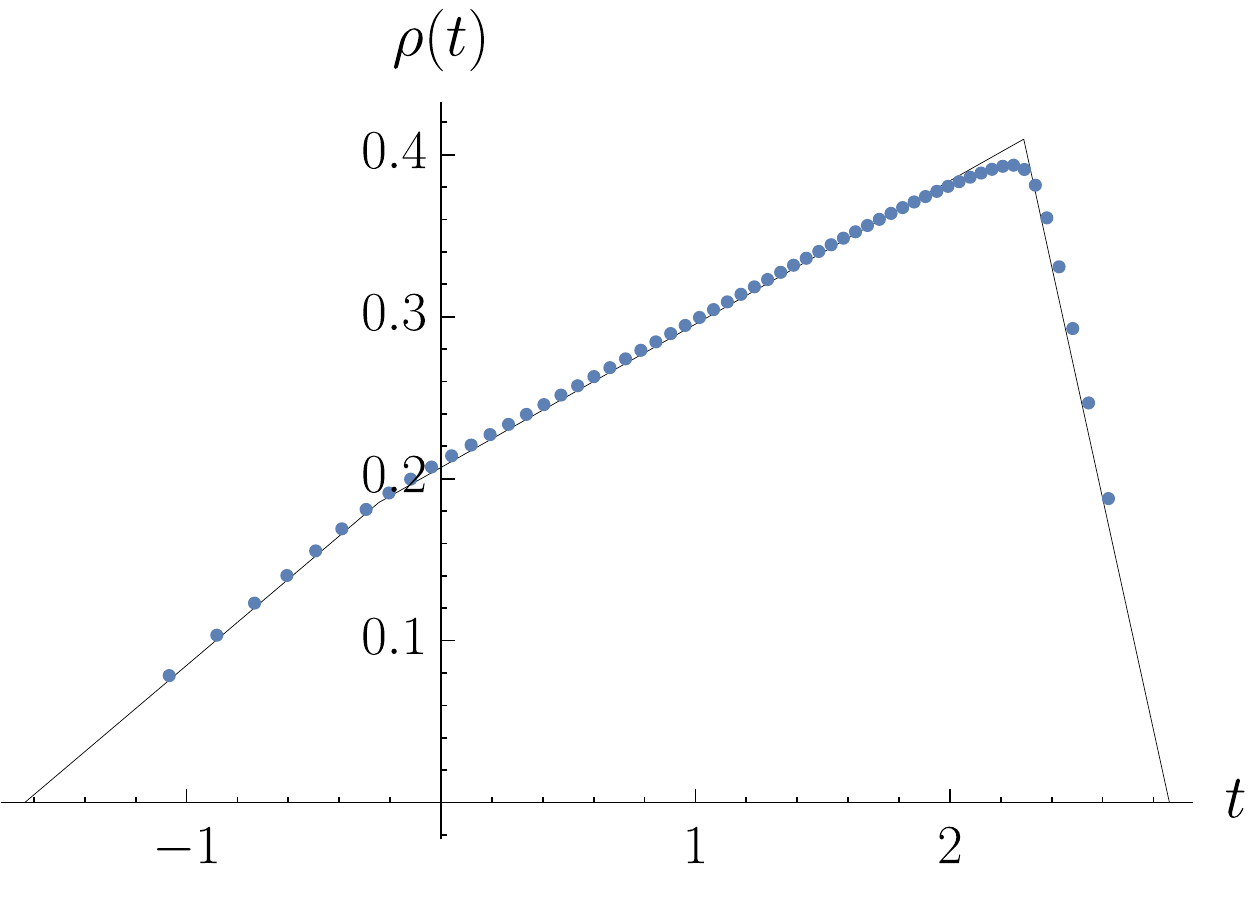}\hss\includegraphics[width=0.48\linewidth]{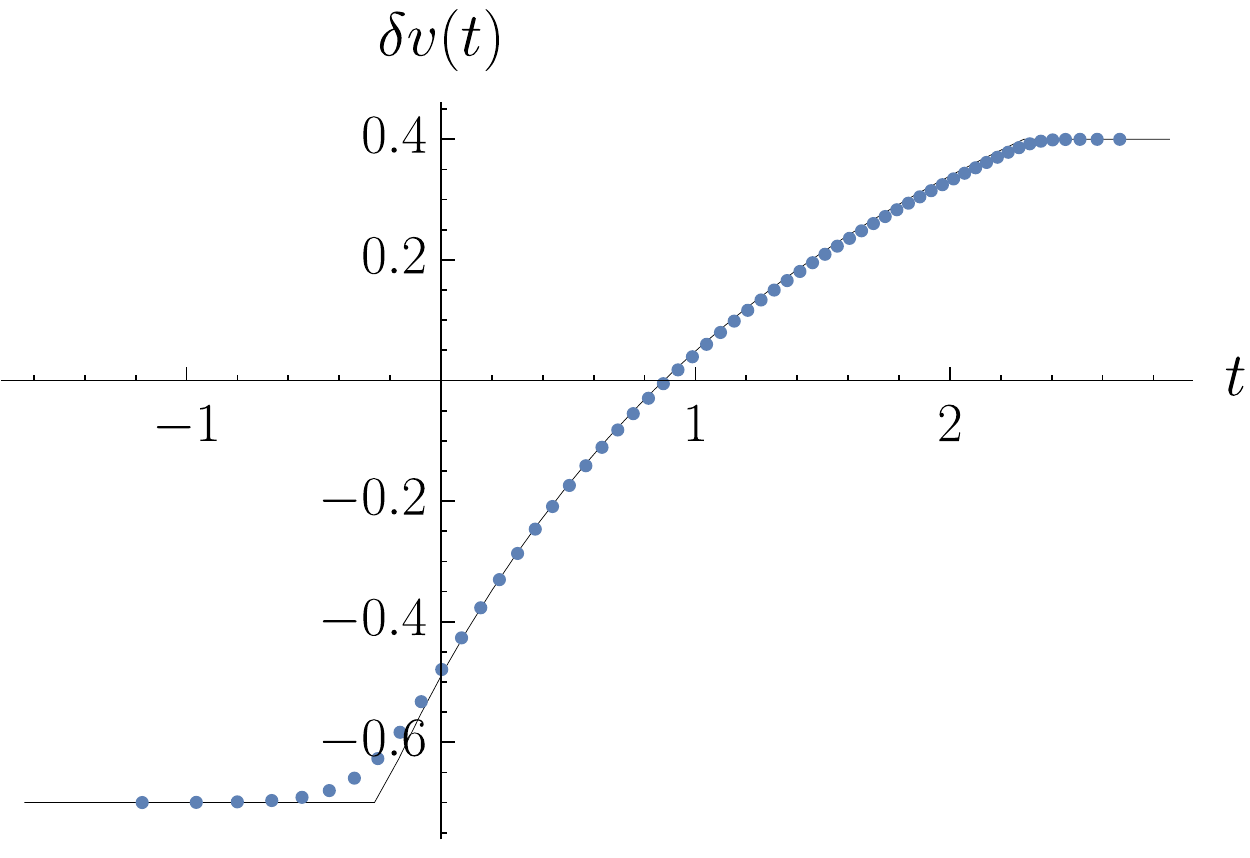}}
\caption{The eigenvalue density $\rho(t)$ and the function $\delta v(t)$ for $\Delta_a=\{0.4,0.5,0.7,2\pi-1.6\}$ and $N=60$, compared with the leading order expression.}
\label{fig:rhodv}
\end{figure}

The main setup is to arrive at a numerical solution to the BAE (\ref{eq:BAE}) through multidimensional root finding using the leading order distribution as the starting point.  We have implemented this in Mathematica using FindRoot.  The solution is first obtained either with MachinePrecision or with WorkingPrecision set to 30, and further refined using WorkingPrecision set to 200 and default settings for AccuracyGoal and PrecisionGoal.  Convergence to a stable solution can be a bit delicate, since the BAE is highly sensitive to the tails; if even a single eigenvalue is sufficiently displaced, then it is easy for FindRoot to fail.  In most cases, we have been able to obtain numerical solutions up to $N\approx200$, although larger values of $N$ are possible with some refinement of the initial distribution.  As an example, the numerical solution for the $u_i$ and $\tilde u_i$ eigenvalues for$\Delta_a=\{0.4, 0.5, 0.7,2\pi-1.6\}$ and $N=60$ is shown in figure~\ref{fig:evals}.  The corresponding eigenvalue density $\rho(t)$ and function $\delta v(t)$ are shown in figure~\ref{fig:rhodv}.  

Once the eigenvalues are obtained, it is then simply a matter of numerically evaluating the index (\ref{eq:logZ}) on the solution to the BAE.   The main challenge here is the evaluation of $\det\mathbb B$, as the Jacobian matrix is ill-conditioned.  (This is why we work to high numerical precision when solving the BAE.)  For a given set of chemical potentials $\Delta_a$, we compute $\log Z$ for a range of $N$.  We then subtract out the leading behavior (\ref{eq:logZ0}) and decompose the residuals into a sum of four independent terms
\begin{equation}
{\rm Re}\log Z={\rm Re}\log Z_0+A+B_1n_1+B_2n_2+B_3n_3,
\end{equation}
where we have used the condition $\sum_a n_a=2$.  At this stage, we then perform a linear least-squares fit of $A$ and $B_a$ to the function
\begin{equation}
f(N)=f_1N^{1/2}+f_2\log N+f_3+f_4N^{-1/2}+f_5N^{-1}+f_6N^{-3/2}.
\end{equation}
We are, of course, mainly interested in $f_2$.  However, since $N$ ranges from about 50 to 200, it is important to consider the first few inverse powers of $N$ as well.  (We have confirmed numerically that the first subdominant term enters at $\mathcal O(N^{1/2})$, and that in particular terms of $\mathcal O(N)$ are absent.)

\begin{table}[t]
\begin{tabular}{l|l|l||l|l|l}
$\Delta_1$&$\Delta_2$&$\Delta_3$&$f_1$&$f_2$&$f_3$\\
\hline
$\pi/2$&$\pi/2$&$\pi/2$&$3.0545$&$-0.4999$&$-3.0466$\\
\hline
$\pi/4$&$\pi/2$&$\pi/4$&$4.2215-0.0491n_1$&$-0.4996+0.0000n_1$&$-4.1710-0.2943n_1$\\
&&&$-0.1473n_2-0.0491n_3$&$+0.0000n_2+0.0000n_3$&$+0.0645n_2-0.2943n_3$\\
\hline
$0.3$&$0.4$&$0.5$&$7.9855-0.2597n_1$&$-0.4994-0.0061n_1$&$-9.8404-0.9312n_1$\\
&&&$-0.5833n_2-0.6411n_3$&$-0.0020n_2-0.0007n_3$&$-0.0293n_2+0.3739n_3$\\
\hline
$0.4$&$0.5$&$0.7$&$6.6696-0.1904n_1$&$-0.4986-0.0016n_1$&$-7.5313-0.6893n_1$\\
&&&$-0.4166n_2-0.4915n_3$&$-0.0008n_2-0.0001n_3$&$-0.1581n_2+0.2767n_3$
\end{tabular}
\caption{Numerical fit for ${\rm Re}\log Z={\rm Re}\log Z_0+f_1N^{1/2}+f_2\log N+f_3+\cdots$.  The values of $N$ used in the fit range from $50$ to $N_{\rm max}$ where $N_{\rm max}=290, 150, 190, 120$ for the four cases, respectively. We made use of the fact that the index is independent of the magnetic fluxes when performing the fit for the special case ($\Delta_a=\{\pi/2,\pi/2,\pi/2,\pi/2\}$).}
\label{tbl:dat}
\end{table}

The results of the numerical fit are presented in Table~\ref{tbl:dat}.  Our main result is that the numerical evidence points to the coefficient of the $\log N$ term being exactly $-1/2$.  We thus have
\begin{equation}
{\rm Re}\log Z=-\frac{N^{3/2}}{3}\sqrt{2\Delta_1\Delta_2\Delta_3\Delta_4}\sum\limits_a\frac{n_a}{\Delta_a}
+N^{1/2}f_1(\Delta_a,n_a)-\fft12\log N+f_3(\Delta_a,n_a)+\mathcal O(N^{-1/2}),
\label{eq:tti}
\end{equation}
where $f_1$ and $f_3$ remain to be determined.  One may wonder whether their dependence on the magnetic fluxes, $n_a$, follows the same leading order behavior, namely $\sum_a n_a/\Delta_a$.  Unfortunately, examination of the table shows that this is not the case.

Although we have been unable to discern the general behavior of $f_1$, for the special case we find the approximate expression
\begin{equation}
f_1=3.0545\approx\fft{11\pi}{8\sqrt{2}}=\fft\pi{\sqrt2}\left(\fft1{24}+\fft13+1\right).
\end{equation}
We have in fact extended the special case to $k>1$.  For $k\sim\mathcal O(1)$, the eigenvalue distribution retains the same features, but with appropriate scaling by $k$.  Working specifically up to $k=5$ and with $N$ up to 200, we find good evidence that in this case the partition function takes the form
\begin{equation}
{\rm Re}\log Z(\Delta_a=\pi/2)=-\fft{\pi\sqrt{2k}}3N^{3/2}+\fft\pi{\sqrt{2k}}\left(\fft{k^2}{24}+\fft13+1\right)N^{1/2}-\fft12\log N+\cdots,
\end{equation}
which may be compared with the ABJM free energy on $S^3$
\begin{equation}
F_{\rm ABJM}=-\fft{\pi\sqrt{2k}}3N^{3/2}+\fft\pi{\sqrt{2k}}\left(\fft{k^2}{24}+\fft13\right)N^{1/2}-\fft14\log N+\cdots.
\end{equation}
While the leading $\mathcal O(N^{3/2})$ term is identical, the first subleading term in the topologically twisted index picks up an additional contribution.  In addition, the coefficients of the log terms differ, and this suggests that the two expressions are capturing distinct features of the holographic dual. Some similarities between the free energy and the topologically twisted index were first pointed out in \cite{Hosseini:2016tor}. More generally, relations between partitions functions on $S^3$ and $S^2\times S^1$ with a topological twist have recently been discussed in  \cite{Closset:2017zgf}. It would be interesting to place our concrete,  subleading in $N$, results within that more formal approach.

\subsection{Perilous $1/N$ expansion}

While the numerical evidence for $-1/2\log N$ appears compelling, ideally this ought to be backed up by an analytical expansion in the large-$N$ limit.  Such an expansion would naturally shed light on the $f_1$ coefficient as well.  However, as mentioned above, the tails make it difficult to maintain a systematic treatment of the $1/N$ expansion.  In particular, the tails occur when the eigenvalues $\{u_i,\tilde u_j\}$ satisfy
\begin{equation}
\tilde u_i-u_i\approx\pm\Delta_a\qquad\Rightarrow\qquad\delta v(t_i)\approx\pm\Delta_a.
\end{equation}
In this case, the logs in the BAE, (\ref{eq:BAE}), for $j$ near $i$ are evaluated near zero.  The resulting large logs cause apparently subleading terms to become important, and hence mixes up orders in the superficial $1/N$ expansion, as already noted in \cite{Benini:2015eyy}.

The leading order partition function may be obtained by properly accounting for the large logs, and we suspect that a careful treatment would allow the computation to be extended to higher orders.  However, this remains a technical challenge, as can be seen from the following illustration.  In the large-$N$ limit, it is natural to focus on the eigenvalue density $\rho(t)$ and the function $\delta v(t)$.  In the formal large-$N$ expansion, both functions are considered to be $\mathcal O(1)$, which is consistent with the plots in figure~\ref{fig:rhodv}.  However, their leading-order slopes are discontinuous where the left and right tails meet the inner interval.  This gives rise to a $\delta$-function divergence when working with their second derivatives.  While the divergence is unimportant at leading order, it presents difficulties at higher order.

Of course, as can be seen in figure~\ref{fig:rhodv}, the actual solution does not have discontinuous slope.  As an estimate, we first note that the range where $\rho(t)$ changes slope is of $\mathcal O(1/\sqrt N)$.  As a result, $\rho''(t)\sim\mathcal O(\sqrt N)$ near the transition points, and a similar estimate can be made for $\delta v(t)$.  While this avoids the $\delta$-function divergences, it nevertheless mixes up orders in the formal large-$N$ expansion.  Furthermore, it is not just the second derivative, but all higher derivatives as well that become important, even when considering just the first subleading correction to the index.

\section{One-loop quantum supergravity}\label{Sec:Gravity}

Based on our numerical evidence, we conjecture that the topologically twisted index has a universal logarithmic correction given by $-1/2\log N$, in contrast with the ABJM free energy that has the factor $-1/4\log N$.  In the latter case, the field theory result was reproduced by a one-loop supergravity computation in \cite{Bhattacharyya:2012ye}.  In particular, the standard AdS$_4$/CFT$_3$ correspondence relates ABJM theory on $S^3$ to M-theory on global AdS$_4\times S^7/\mathbb Z_k$ \cite{Aharony:2008ug}.  The logarithmic term then originates purely from a ghost two-form zero mode contribution on AdS$_4$.

In the present case, however, we take ABJM theory on $S^2\times S^1$ with a topological twist generated by background magnetic flux.  This topological twist relevantly deforms the ABJM theory to flow toward a superconformal quantum mechanics on $S^1$. Holographically, such an RG flow can be thought of as a Euclidean asymptotically AdS$_4$ BPS magnetic black hole, interpolating between the asymptotically AdS$_4$ region and an AdS$_2\times$ S$^2$ near horizon region. The solution can be embedded into 11-dimensional supergravity \cite{Cvetic:1999xp}, and such an embedding makes it also natural to consider the quantum correction from an 11-dimensional point of view. 

We are thus interested in computing the one-loop correction to the supersymmetric partition function in the BPS black hole background that interpolates between asymptotic AdS$_4\times S^7$ and AdS$_2\times M_9$ near the horizon, where $M_9$ is a $S^7$ bundle over $S^2$.  As a simplification, however, we assume a decoupling limit exists, so that we can focus mainly on the AdS$_2\times M_9$ near horizon geometry.  Alternatively, corrections to the black hole entropy may be considered via the quantum entropy function in the near horizon geometry proposed in \cite{Sen:2008vm}. For extremal black hole with no electric charge, the quantum entropy function reduces to the partition function of 11-dimensional supergravity compactified in the near horizon geometry, and we are again led to AdS$_2\times M_9$.

In the computation of one-loop corrections to the partition function, we focus on the logarithmic term, as such a term, in odd dimensional spaces, arises purely from zero modes  (see \cite{Vassilevich:2003xt} and \cite{Bhattacharyya:2012ye} for a review). The effect of zero modes on the logarithmic term can be naturally divided into two parts: the subtraction of zero modes from the trace of the heat kernel to make the heat kernel well defined, and the integration over zero modes in the path integral. Those two parts can be summarized schematically, for a given kinetic operator $D$ of a physical field, as 
\begin{equation}
\Delta F_D= (-1)^D(\beta_D-1)n_D^0\log L, 
\end{equation}
where $\beta_D$ encodes the integration over zero modes in the path integral, and $-1$ is due to the subtraction in the heat kernel. We use $(-1)^D$ to distinguish bosonic/fermionic contributions. The treatment for ghosts is slightly different, and they are considered separately as in \cite{Bhattacharyya:2012ye}.  In summary, the total logarithmic correction is given by
\begin{equation}
\Delta F= \sum_{\{D\}}(-1)^D(\beta_D-1)n_D^0\log L+ \Delta F_{\mathrm{Ghost}},
\end{equation}
where the summation is over physical fields.

For completeness, we shall first summarize the fields that have non-trivial zero modes $n_D^0$ in AdS$_2$ and their $\beta_D$, although they are quite standard and well known in the literature  (see for example, the appendix of \cite{Sen:2012cj}). We then compute the logarithmic correction from the physical sector and the ghost sector of 11-dimensional supergravity in the near horizon geometry AdS$_2\times M_9$.

\subsection{The number and scalings of zero modes}

The spectrum of a kinetic operator on a non-compact space, such as AdS$_2$, typically consists of two parts: a continuous part due to the non-compactness of the space, and possibly a discrete part that contains a countably infinite number of eigenfunctions with zero eigenvalue. The continuous part of the trace of the heat kernel in the case of AdS$_N$ is well defined, whereas the zero modes from the discrete part, if any, should be subtracted from the heat kernel. The formal sum that counts the number of zero modes in the compact case is divergent when the space is non-compact, 
\begin{equation}\label{n0}
n_0=\sum_j \int \sqrt{g}d^2x |\phi_j(x)|^2, 
\end{equation}
where $\phi_j(x)$'s are normalized to 1. Thus computing $n_0$ requires regularization. For symmetric spaces $G/H$,  $n_0$ can be evaluated by working out explicit eigenfunctions, exchanging the sum and integral, and using a regularized volume as in \cite{Sen:2012cj,Bhattacharyya:2012ye}.

Here, we present another way of computing $n_0$ using the general theorem in \cite{CAMPORESI199457}.  The number of zero modes can be associated with the formal degree of the discrete series representation of $G$ corresponding to the given field, which occurs when $G$ has a maximal torus that is compact. For AdS$_N=SO(N,1)/SO(N)$, they occur when $N$ is even, and they can be labeled in terms of the highest weight label $(\sigma, n_0)$, where $\sigma=(n_\frac{N-2}{2}, n_\frac{N-4}{2}, \dots, n_1)$, with $n_\frac{N-2}{2}> n_\frac{N-4}{2}>\dots > n_1>|n_0|$.  Any vector bundle over AdS$_N$ can be labeled by an irreducible representation of $SO(N)$ (or $\text{Spin}(N)$) in terms of highest weight labels $\tau=(h_\frac{N}{2}, h_\frac{N-2}{2}, \dots, h_1)$, and in order to determine the number of zero modes for a given field, one looks for the branching condition
\begin{equation}\label{branching}
\frac{1}{2}<|n_0|\leq|h_1|\leq n_1 \leq \dots \leq n_{\frac{N-2}{2}} \leq h_{\frac{N}{2}}.
\end{equation}
The number of zero modes is the sum of all degrees $P(\sigma, n_0)$ of discrete series representations $(\sigma, n_0)$ that satisfies the branching condition, up to a normalization factor that only depends on the dimension: 
\begin{equation}\label{zeromodes}
n_0^{\tau}=\frac{\text{Vol}(\mathrm{AdS}_N)}{c_N}\sum_{(\sigma,n_0)} P(\sigma, n_0).
\end{equation}
For AdS$_2$, $P(n_0)=n_0-\frac{1}{2}$ and $c_N=2\pi$, and a field is labeled by a single highest weight label which is its spin.  (General expressions for $c_{N}$ and $P(\sigma,n_0)$ can be found in section~6 of \cite{CAMPORESI199457}.) The branching condition, (\ref{branching}), implies that fields with spin greater than $\frac{1}{2}$ have zero modes, i.e. one-form, gravitino, and graviton fields. Moreover, using (\ref{zeromodes}), one has 
\begin{align}
n^0_{g}&=2\times \frac{(-2\pi)}{2\pi}\left(2-\frac{1}{2}\right)=-3,\nn\\
n^0_{\psi}&=2\times \frac{(-2\pi)}{2\pi}\left(\frac{3}{2}-\frac{1}{2}\right)=-2,\nn\\
n^0_A&=2\times \frac{(-2\pi)}{2\pi}\left(1-\frac{1}{2}\right)=-1,
\end{align}
where $n^0_{g}$, $n^0_{\psi}$, $n^0_{A}$ are respectively the number of zero modes of a graviton, a gravitino and a one form. We also used the fact that the regularized volume of AdS$_2$ is $-2\pi$. These values, of course, coincide with the direct evaluation performed in \cite{Sen:2012cj}.

The logarithmic part of the integration over zero modes in the path integral can be obtained by dimensional analysis.  Given a kinetic operator $\mathcal{O}$, the path integral over zero modes is given by 
\begin{equation}\label{intzeromodes}
\int Df|_{\text{zero modes}} \exp\left(-\int d^dx\sqrt{g} f \mathcal{O} f\right)=\int Df|_{\text{zero modes}}\sim L^{\beta_{\mathcal{O}}n_{\mathcal{O}}^0},
\end{equation}
through which we define $\beta_{\mathcal{O}}$ for an operator $\mathcal{O}$. To obtain the logarithmic correction, it is enough to find the $L$ dependence of (\ref{intzeromodes}), which amounts to finding the $L$ dependence in the path integral measure. In the case of Euclidean AdS$_{2N}$, all such zero modes arise due to a non-normalizable gauge parameter $\lambda$, where $f=G\lambda$ with $G$ representing the infinitesimal gauge transformation.  For example, let $g_{\mu\nu}=L^2g^{(0)}_{\mu\nu}$. The path integral measure of a $p$-form in $d$ dimensions is normalized as
\begin{equation}
\int D A_{[p]} \exp\left(-L^{d-2p}\int d^dx \sqrt{g^{(0)}} g^{(0)\mu_1\nu_1 }g^{(0)\mu_2\nu_2 }\dots g^{(0)\mu_p\nu_p }A_{\mu_1\dots \mu_p}A_{\nu_1\dots \nu_p}\right)=1.
\end{equation}
Therefore, the correctly normalized measure is 
\begin{equation}
D(L^{\frac{d-2p}{2}}d \lambda_{[p-1]}),
\end{equation}
where $\lambda_{[p-1]}$ is a non-normalizable $(p-1)$-form gauge parameter, and has no $L$ dependence. Such a measure gives $L^{(d-2p)/{2}}$ per zero mode, and therefore contributes as $L^{(d-2p)n_p^0/{2}}$ in the path integral. Thus $\beta_{A_{[p]}}=(d-2p)/{2}$ in $d$ dimensions. One can carry out similar computations for other fields, paying particular attention to the possible $L$ dependence of the gauge parameter, as in \cite{Sen:2012cj} and \cite{Banerjee:2010qc}. One then finds
\begin{equation}
\beta_g= \frac{D}{2}, \qquad \beta_{\psi_{\mu}} = D-1, \qquad \beta_{A_{[p]}}=\frac{d-2p}{2}.
\end{equation}

\subsection{The logarithmic corrections}
The 11-dimensional $\mathcal{N}=1$ gravitational multiplet consists of ($g_{\mu\nu}, \psi_{\mu}, C_{\mu\nu\rho}$). The fluctuation of the metric to the lowest order can be summarized as 
\be
h_{\mu\nu}(x,y)=
\begin{cases}
h_{\alpha \beta}(x)\phi(y), \\
h_{\alpha i}=\sum_a A_{\alpha}^a(x)K^a_{i}(y),\\
\phi(x)h_{ij}(y),
\end{cases}
\ee
where we use $(x^{\alpha}, y^i)$ to denote AdS$_2$ and $M_9$ coordinates, respectively, and $K^{a i}(y)\partial_i$ is a killing vector of $M_9$. The graviton zero modes therefore contribute in two ways: a graviton in AdS$_2$, and gauge fields corresponding to Killing vectors of $M_9$.

From the near horizon geometry in \cite{Benini:2015eyy} one can read off the metric on $M_9$
\begin{equation}\label{nearhorizon}
    ds^2_9=\Delta^{\frac{2}{3}}ds^2_{S^2}+\frac{4}{\Delta^{\frac{1}{3}}}\sum_{i=1}^4 \frac{1}{X_i}\left(d\mu_i^2+\mu_i^2(d\psi_i+\frac{n_i}{2}\cos\theta d\phi)^2\right),
\end{equation}
where we denote the coordinates on $S^2$ by $(\theta, \phi)$, $X_i$'s are constant with $\prod X_i=1$, $\Delta=\sum_{i=1}^4X_i \mu_i^2$, and $\sum_{i=1}^4\mu_i^2=1$. The metric, (\ref{nearhorizon}), suggests the following seven Killing vectors:
\begin{align}
&\Bigl\{\cos\phi\partial_{\theta}-\cot\theta \sin \phi \partial_{\phi}+\sum_{j}\frac{n_j}{2}\frac{\sin \phi}{\sin \theta}\partial_{\psi_j},~
-\sin\phi \partial_{\theta}-\cot\theta\cos\phi\partial_{\phi}+\sum_j \frac{n_j}{2}\frac{\cos\phi}{\sin\theta }\partial_{\psi_j},~
\partial_{\phi}\Bigr\},\nn\\
&\Bigl\{\partial_{\psi_i}\Bigr\},
\end{align}
where $i=1,2,3,4$, and the Killing vectors span the algebra of the isometry group $SU(2)\times U(1)^4$. Thus the logarithmic correction due to the 11-dimensional graviton is given by 
\begin{equation}\label{graviton}
\Delta F_h= (\beta_h-1)(n_g^0+7n_{A}^0)\log L= \left(\frac{11}{2}-1\right)[(-3) \times 1 + (-1)\times 7]\log L=-45 \log L.
\end{equation}

A gravitino $\psi_{\mu}$ can either be an AdS$_2$ gravitino and a spin-1/2 fermion on $M_9$, or vice versa. Ideally one would find the number of killing spinors of $M_9$. Nevertheless, it is more convenient to reduce to four-dimensions first.  In this case, the $\mathcal{N}=2$ gravitational multiplet contains two gravitinos, which further decompose to two gravitinos on AdS$_2$. As the number of gravitinos only concerns the number of supersymmetries that are preserved, it should be the same no matter whether one works directly in 11 dimensions, or through a reduction to four dimensions.  Thus, the contribution due to the gravitino is given by 
\begin{equation}\label{gravitino}
\Delta F_{\psi}=-\sum (\beta_{\psi}-1)n_{\psi}^0\log L=-(10-1)[(-2)\times 2]\log L=36\log L, 
\end{equation}
where the minus sign is assigned as it is Grassmann odd.  

The fluctuation of a 11 dimensional 3 form can be summarized as
\be\label{3form}
C_3(x,y)=
\begin{cases}
B_{3}(y),\\
A_{1}(x)\wedge B_2(y),\\
A_{2}(x)\wedge B_1(y),
\end{cases}
\ee
where the subscript represents the rank of the form, $A(x)$ represents a form on AdS$_2$ and $B(y)$ a form on $M_9$. Note for $M_9$ the Betti numbers $b_1=0$ and $b_2=1$.
Therefore the contribution from the 3-form, from the first line in (\ref{3form}), is 
\begin{equation}\label{threeform}
\Delta F_{C}=(\beta_C-1)n^0_C\log L=\left (\frac{5}{2}-1\right)[ (-1)\times 1]\log L=-\frac{3}{2} \log L. 
\end{equation}

We now turn to the treatment for ghosts, which requires special care. We therefore compute them separately, and we only concern ourselves with ghosts that give rise to AdS$_2$ zero modes.  Therefore only the ghosts for the graviton, which gives a vector ghost $c_\mu$, and the ghosts for the 3-form are considered. The BRST quantization of supergravity generally provides a kinetic term $c^*_\mu(-g^{\mu\nu}\Box -R^{\mu\nu})c_\nu$ with other off diagonal terms that are lower triangular, which do not change the eigenvalues of the kinetic operator on $c_\nu$. In our case, $R_{\mu\nu}$ is never zero, and therefore the graviton ghosts are not relevant to the logarithmic correction. 

The general action for quantizing a $p$-form $A_p$ requires a set of $(p-j+1)$-form ghost fields, with $j=2, 3, \dots, p+1$, and the ghost is Grassmann even if $j$ is odd and Grassmann odd if $j$ is even \cite{Siegel:1980jj,Copeland:1984qk}. Although for the $(p-j+1)$-form, the Laplacian operator $(\Delta_{p-j+1})^{j}$ in the computation of the heat kernel requires an extra $j-1$ removal of the zero modes, the integration over the zero modes is unchanged. That results, as in Eq.~(3.4) of \cite{Bhattacharyya:2012ye}, is
\begin{equation}
\Delta F_{\text{Ghost}}=\sum_j (-1)^j(\beta_{A_{p-j}}-j-1)n_{A_{p-j}}^0\log L.
\end{equation}
Note for our case that $b_1$ of $M_9$ is zero. Therefore the only non-vanishing term is $p=3$, $j=2$, which gives 
\begin{equation}\label{ghost}
\Delta F_{\text{Ghost}}=-\frac{3}{2}\log L.
\end{equation}

Finally, adding the contributions (\ref{graviton}), (\ref{gravitino}), (\ref{threeform}) and (\ref{ghost}) leads to the total logarithmic correction 
\be\label{result}
\Delta F=\left(-45+36-\frac{3}{2}-\frac{3}{2}\right)\log L=-12\log L\sim -2\log N,
\ee
where in the last equality we used the AdS/CFT dictionary $N\sim L^6$, and neglected $L$ independent terms.  We note that this result does not match with the logarithmic term of the topologically twisted index, (\ref{eq:tti}), which instead has coefficient $-1/2$.

We finish this section by addressing a very natural question. In our computation we have focused exclusively on the near horizon geometry.  Given that the black holes we are discussing are asymptotically AdS$_4$, are there contributions that come precisely from the asymptotic region? After all, the computation of \cite{Bhattacharyya:2012ye} obtained logarithmic corrections on the gravity side by studying quantum supergravity on AdS$_4\times S^7$ and found that the entire contribution comes from a two-form zero mode in AdS$_4$. The result of \cite{Bhattacharyya:2012ye} perfectly matches field theory results corresponding to the free energy of ABJM on $S^3$.  Our case, however, pertains to a computation of ABJM on $S^2\times S^1$.  In an elucidating discussion about boundary modes presented in \cite{Larsen:2015aia}, the authors considered global aspects of  AdS$_4$ with $S^3$ and $S^2\times S^1$ boundary conditions. In particular, they established that the Euler number depends on these boundary conditions and is, respectively, $\chi=1$ and $\chi=0$. This result indicates the existence of a two-form zero mode in the case of $S^3$  boundary conditions which is precisely the two-form responsible for the successful match with the field theory free energy. It also indicates the absence of the corresponding two-form zero mode for $S^2\times S^1$ boundary conditions. Moreover, the crucial use of $S^3$ boundary conditions in the explicit construction of the non-trivial two forms \cite{CAMPORESI199457,10.2307/2042193,Donnelly1981}, also supports our claim. 

Therefore, at least to this level of scrutiny, there is no contribution coming from the asymptotically AdS$_4$ region.  It will, of course, be interesting to develop a systematic approach to dealing with asymptotically AdS contributions in the framework of holographic renormalization. 
 
\section{Discussion}\label{Sec:Conclusions}

Given the disagreement in the computations, we shall discuss some of our underlying assumptions.  On the field theory side, the topologically twisted index reproduces the Bekenstein-Hawking entropy of AdS black holes at leading order in the large-$N$ expansion \cite{Benini:2015eyy,Cabo-Bizet:2017jsl}.  It is thus tempting to expect that the index provides an complete microstate description at all orders.  To explore this possibility, we have performed a numerical investigation of the topologically twisted index and obtained a logarithmic correction of $-1/2\log N$.  It is this term that we have attempted to reproduce by computing a one-loop partition function on the supergravity side of the duality.

While AdS/CFT suggests that the corresponding one-loop partition function ought to be computed in the full magnetic AdS$_4$ black hole background, we made a decoupling approximation and focused instead on the AdS$_2\times S^2$ near horizon region.  Given the 11-dimensional supergravity origin, only zero modes contribute to the logarithmic term, and we find instead the term $-2\log N$ from the bulk computation.  Of course, this treatment of separate near horizon and asymptotic regions is somewhat {\it ad hoc}, and it is entirely possible that agreement would be restored if we instead worked in the full geometry.  Although one may expect the degrees of freedom responsible for the entropy to be close to the horizon, it would be important to perform a more careful investigation to see whether or not this is truly the case.

Of course, this all assumes that the asymptotically AdS$_4$ black hole partition function is the correct object to match with the topologically twisted index.  An alternative framework is to compare with the quantum entropy function \cite{Sen:2008yk,Sen:2008vm}.  However, this approach is technically similar and gives the same result of $-2\log N$.  In this case, the disagreement suggests that the quantum entropy formalism needs to be modified for the case of asymptotically AdS black holes.

Another point that deserves discussion is the role of extremization of the index in order to match with the entropy on the gravity side \cite{Benini:2015eyy}. Our field theory computation was performed for arbitrary magnetic charges and potentials. Extremization, at leading order, relates the potentials, $\Delta_a$, to the magnetic charges $n_a$. We believe that the $\log N$ coefficient does not change at the level that we discuss because the leading order relation from extremization is independent of $N$. 

More generally, the field theory index is clearly grand canonical while the gravity side is microcanonical. One can think of the extremization procedure of \cite{Benini:2015eyy} as performing a Legendre transform where the main contribution comes from electrically neutral configurations.  Although it is plausible that under certain conditions a partition function equals the entropy, the equivalence of the topologically twisted index with the entropy of magnetically charged black holes deserves a more systematic discussion within the framework of holographic renormalization. 

Let us now  discuss a number of other directions that would be nice to explore. One natural question is motivated by the universality of the result of \cite{Bhattacharyya:2012ye}. Indeed, a large class of field theory partition functions on $S^3$  has a  $-1/4\log N$ correction for matter Chern-Simons theories of various types \cite{Fuji:2011km,Marino:2011eh}. On the gravity side of the correspondence, the universality of this result relies on the logarithmic term being given strictly by a two-form zero mode in AdS$_4$; it is thus independent of the Sasaki-Einstein $X_7$ manifold where the supergravity is defined \cite{Bhattacharyya:2012ye}. Despite the current disagreement, it would be interesting to entertain a similar universality argument for the correction we find here, namely $-1/2\log N$.  Although we do not fully understand the gravity side, it is clear that the answer we provide relies on very general properties of $M_9$ such as the number of Killing vectors and the second Betti number. The underlying question behind this direction rests on the various hints    \cite{Hosseini:2016tor,Closset:2017zgf} for connections between the partition function on $S^3$ and the topologically twisted index in $S^2\times S^1$ that we have now investigated beyond the large-$N$ limit.

Although we have proposed an {\it ad hoc} treatment for the asymptotically AdS$_4$ region, the more general question of how to systematically apply the principles of holographic renormalization for the computation of logarithmic corrections to black hole entropy remains open. A better understanding, perhaps including generalizations of the quantum entropy function, is clearly needed to tackle the plethora of solutions that are intrinsically four-dimensional, and in particular, more general solutions of ${\cal N}=2$ gauged supergravity. 

A more challenging question is:  Can one obtain the full logarithmic correction to the entropy, and not just the $\log N$ coefficient? One possibility is to tackle the theory directly in four dimensions. In this case the heat kernel, being in an even dimensional space, contributes in a more complicated way. A similar technical problem appears in the 't~Hooft limit where the gravity dual theory lives on AdS$_4\times \mathbb{CP}^3$. It is worth pointing out an added difficulty in the case of the magnetically charged black holes we are considering.  For asymptotically flat black holes, a typical practice is to consider particular $N$-correlated scalings  of the charges; this allows for the computation of corrections in various regimes. However, generic scalings of the charges are not allowed in our case because the charges are constrained, for example, by $\sum n_i=2$. Alternatively, one could attempt a full supergravity localization following the work \cite{Dabholkar:2014wpa} and the more recent effort in \cite{Nian:2017hac}.

Of course, it is worth noting that the first subleading correction to the topologically twisted index occurs at $\mathcal O(N^{1/2})$.  In principle, it would be useful to obtain an analytic expression for this correction, which we denoted $f_1(\Delta_a,n_a)$ in (\ref{eq:ABJMrlz}).  On the gravity side, this term presumably originates from higher-derivative corrections to the Wald entropy.  While we have been as yet unable to find the analytic form of $f_1$, it may be possible to do so with additional numerical work.

Finally, it would be interesting to discuss other asymptotically AdS gravity configurations forming AdS/CFT dual pairs. For example, we may consider black strings in AdS$_5$ that are dual to topologically twisted four-dimensional field theories \cite{Benini:2013cda}. The topologically twisted index for the dual four-dimensional field theories on $S^2\times T^2$ has been constructed in \cite{Benini:2016hjo,Honda:2015yha} and its high temperature limit has recently been discussed in \cite{Hosseini:2016cyf}. It seems also possible, with the new insight provided recently in  \cite{Hosseini:2017mds}, to return to the question of microstate counting for the asymptotically AdS$_5$ black holes more generally.

\acknowledgments
We are thankful to F. Benini, A. Cabo-Bizet, A. Charles, C. Closset, A. Gnecchi, A. Grassi, S. M. Hosseini, K. Hristov, U. Kol, F. Larsen, M. Mari\~no, I. Papadimitriou, G. Silva  and S. Vandoren for various discussions on closely related topics. VR and LPZ thank the ICTP   and CERN, respectively, for warm hospitality. LPZ is particularly thankful to Ashoke Sen for various generous discussions on the subject of logarithmic corrections to black hole entropy.  While we were completing this work, we became aware that R.\ Gupta, I.\ Jeon and S.\ Lal have independently obtained the logarithmic correction from the QEF.  This work is partially supported by the US Department of Energy under Grant No.\ DE-SC0007859 and No.\ DE-SC0017808.

\bibliographystyle{JHEP}
\bibliography{BHLocalization}

\end{document}